\documentclass{article}
\usepackage{graphicx}
\graphicspath{%
    {converted_graphics/}
    {C:/Dropbox/1-kgl-top/Papers/2014/EIT-ParticleDynamics/}
}
\begin{document}

\begin{center}
{\huge Particle Dynamics in Damped Nonlinear}\vskip8pt

{\huge Quadrupole Ion Traps}\vskip12pt

{\Large Eugene A. Vinitsky, Eric D. Black, and Kenneth G. Libbrecht}\vskip4pt

{\large Department of Physics, California Institute of Technology}\vskip-1pt

{\large Pasadena, California 91125}\vskip-1pt

\vskip18pt

\hrule\vskip1pt \hrule\vskip14pt
\end{center}

\textbf{Abstract.} We examine the motions of particles in quadrupole ion
traps as a function of damping and trapping forces, including cases where
nonlinear damping or nonlinearities in the electric field geometry play
significant roles. In the absence of nonlinearities, particles are either
damped to the trap center or ejected, while their addition brings about a
rich spectrum of stable closed particle trajectories. In three-dimensional
(3D) quadrupole traps, the extended orbits are typically confined to the
trap axis, and for this case we present a 1D analysis of the relevant
equation of motion. We follow this with an analysis of 2D quadrupole traps
that frequently show diamond-shaped closed orbits. For both the 1D and 2D
cases we present experimental observations of the calculated trajectories in
microparticle ion traps. We also report the discovery of a new collective
behavior in damped 2D microparticle ion traps, where particles spontaneously
assemble into a remarkable knot of overlapping, corotating diamond orbits,
self-stabilized by air currents arising from the particle motion.

\section{Introduction}

Electrodynamic ion traps, also known as Paul traps or quadrupole ion traps
(QITs), have found a broad range of applications in physics and chemistry,
including precision mass spectrometry \cite{spect}, quantum computing \cite%
{blatt}, and improved atomic frequency standards \cite{wineland}. When
trapping atomic or molecular ions, QITs often operate with radiofrequency
electric fields in vacuum, and the particle dynamics in these traps have
been well studied over many decades \cite{spect, iontrapbook, march}.

Microparticle quadrupole ion traps (MQITs) are also commonly used to measure
the detailed properties of individual charged particles in the 100 nm to 100 
$\mu $m size range, including aerosols \cite{aerosol1, aerosolspect}, liquid
droplets \cite{kramer, arnold}, solid particles \cite{wuerker, visan,
schlemmer}, nanoparticles \cite{cai, snit}, and even microorganisms \cite%
{microorganisms, microorganisms2}. MQITs have also become popular in physics
teaching, as they provide a fascinating and somewhat counterintuitive
demonstration of oscillatory mechanics and electric forces. Moreover, MQITs
are quite inexpensive to construct, levitating particles in air using 50-60
Hz electric fields, making them well suited for teaching \cite{winter,
robertson2, wuerker}. Besides individual particles, MQITs can also trap
large numbers of charged particles that self-organize into Coulomb
crystalline structures \cite{robertson2, wuerker, coulomb}.

The addition of motional damping, for example from gas damping or laser
cooling, significantly alters the particle dynamics in QITs, and there has
been considerable interest in understanding these effects \cite{whitten,
foot, whetten, hasegawa, ziaetan2}, particularly in spectroscopic
applications when weak damping is used to stabilize the particle motions 
\cite{march}. For the simplest and best-studied case -- linear damping in
purely quadrupolar electric fields -- the addition of damping enlarges and
shifts the stability regions in parameter space \cite{foot}, but the
boundaries between stable and unstable regions remain sharp. In other words,
particles either spiral down until they are at rest at the trap center, or
the oscillating electric forces overpower the damping and eject particles
from the trap. Thus for the simplest damped QITs, the only stable dynamical
solution to the trap equations is the $\vec{x}(t)=0$ solution.

We have found, however, that the situation changes markedly with the
addition of weak nonlinearities in the trap equations -- either
nonlinearities in the field geometry (a deviation from a pure quadrupole
field configuration) or nonlinearities in the damping. In both cases the
boundaries between stable and unstable regions in parameter space may no
longer be sharp, and nontrivial stable solutions to the trap equations
appear. We use the term \textquotedblleft extended orbits\textquotedblright\
for these solutions, describing particles that execute stable, closed
oscillatory trajectories within a damped ion trap.

We have calculated and experimentally confirmed several examples of these
extended orbits, as described below. Nonlinear field geometries in QITs have
been investigated by several authors \cite{ziaetan, blumel, blumel2,
mihalcea, alheit}, but to our knowledge the different types of extended
orbits in damped nonlinear QITs have not previously been characterized. We
have found that these states appear quite readily in MQITs when the drive
voltage is sufficiently high. After observing this behavior frequently in
our own laboratory investigations, we also identified similar examples in
online videos \cite{vid1, vid2}. Although it appears that extended orbital
behaviors are fairly common in MQITs, we were not able to find an adequate
explanation of these observations in our literature search.

The detailed characteristics of an extended orbit depends on the trapped
particle properties, including its charge, mass, and radius, so the
appearance of a specific dynamical behavior could serve as a convenient and
accurate measurement tool in MQITs. And since the extended orbits are
stable, measurements of the orbital properties are nondestructive in that
they do not eject particles from the trap. Although we have identified
several examples of extended orbital behavior, the parameter space of
nonlinearities in trap geometry is large, so additional examples may exist.
Whether any of these novel dynamical behaviors can be gainfully harnessed in
ion trapping applications remains a question for additional study.

\section{Axial Motion in 3D Damped Ion Traps}

To connect to the existing ion-trapping literature, we begin with a
description of the equations of motion for a single charged particle in a
purely quadrupolar field geometry with the simplest linear damping,
following the standard notation \cite{whitten, march}. In particular, we
consider a 3D quadrupole trap in cylindrical $(r,z)$ coordinates \cite{march}%
, focusing on the equation of motion describing the axial motion $z(t).$ In
our damped 3D MQITs, the extended orbits we have observed were all confined
to the $z$ axis. The radial trapping forces, including damping, keep the
particle confined to $r=0$ even in the presence of extended motions in $z$,
reducing the dynamical problem to one dimension. Although the 1D equation of
motion cannot fully describe all aspects of particle motion in a 3D damped
ion trap \cite{foot}, we have nevertheless found that it captures the
essence of the extended orbits we have observed. We therefore begin with the
the simpler 1D problem as a means of characterizing these extended orbital
motions.

We write the particle equation of motion 
\begin{equation}
m\frac{d^{2}z}{dt^{2}}+\gamma \frac{dz}{dt}=QE_{z}(z,t)
\label{equationofmotion}
\end{equation}%
where $z$ is axial position in the trap, $m$ is particle mass, $\gamma $ is
the linear damping coefficient, $Q$ is the particle charge, and $E_{z}$ is
the axial electric field. Including AC and DC quadrupole electric fields
then gives the standard trap equation 
\begin{equation}
\frac{d^{2}z}{d\xi ^{2}}+b\frac{dz}{d\xi }+\left[ a-2q\cos \left( 2\xi
\right) \right] z=0  \label{dimensionlessequation}
\end{equation}%
where $\xi =\Omega t/2$ is the dimensionless time, $b=2\gamma /m\Omega $ is
the damping parameter, $a=-4QA_{DC}/m\Omega ^{2}$ is the DC electric force
parameter, $E_{DC}=A_{DC}z$ is the DC electric field, $q=2QA_{AC}/m\Omega
^{2}$ is the AC force parameter, and $E_{AC}=A_{AC}z\cos \left( 2\xi \right) 
$ is the AC electric field.

Previous treatments of damped QITs in the absence of nonlinearities (in
either the damping or the electric field geometry) \cite{whitten, foot,
whetten, hasegawa} have shown that it is possible to eliminate the $dz/d\xi $
term by substituting $z=u\exp \left( -b\xi /2\right) ,$ giving 
\begin{equation}
\frac{d^{2}u}{d\xi ^{2}}+\left[ \tilde{a}-2q\cos \left( 2\xi \right) \right]
u=0
\end{equation}%
where $\tilde{a}=a-b^{2}/4.$ This equation has the form of the Mathieu
equation, which has a well-studied stability diagram \cite{march,
iontrapbook}$.$ As is described in \cite{whitten, foot}, stability in $z$ is
different from stability in $u,$ owing to the additional $\exp \left( -b\xi
/2\right) $ factor. In \cite{foot} the authors plot stability diagrams in
the $(a,q)$ plane for several nonzero $b$ values, showing that the stable
regions are larger and shifted relative to the corresponding regions when $%
b=0.$ As described in these references, particles decay to $z=0$ within the
stable regions of parameter space, and are expelled from the trap outside
the stable regions.

In our experiments with MQITs, the particles are large enough that the
gravitational force is significant, plus we often add a uniform electric
field $E_{0}$ that produces a constant force similar to gravity. With both
forces in the $z$ direction, this adds an additional downward force $%
mg_{eff}=(mg+QE_{0})$ to Equation \ref{equationofmotion}. After
transformations and focusing on the $a=0$ special case, Equation \ref%
{dimensionlessequation} becomes 
\begin{equation}
\frac{d^{2}\tilde{z}}{d\xi ^{2}}+b\frac{d\tilde{z}}{d\xi }-2q\tilde{z}\cos
\left( 2\xi \right) =-1  \label{withgravity}
\end{equation}%
where $\tilde{z}=z\Omega ^{2}/4g_{eff},$ and the other parameters are the
same as above.

\begin{figure}[htb] 
  \centering
  \includegraphics[height=3.5in,keepaspectratio]{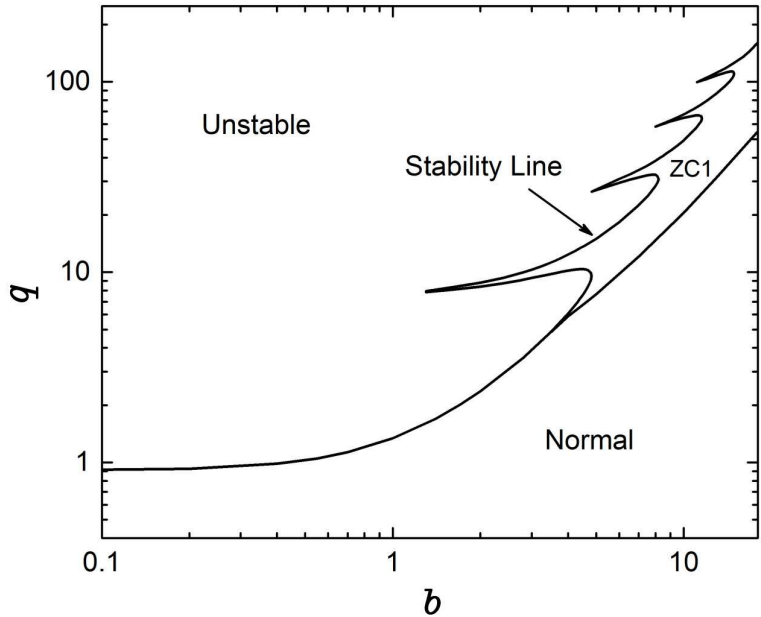}
  \caption{Stability diagram in $(b,q)$
describing solutions to Equation \protect\ref{withgravity}. Above the
labeled stability line, particles are ejected from the trap. Below the
stability line, in the \textquotedblleft Normal\textquotedblright\ region,
particles exhibit a simple oscillatory behavior with $\tilde{z}<0$ for all $%
\protect\xi .$ In the ZC1 region, particles exhibit extended zero-crossing
orbits. Examples of both these behaviors are shown in Figure \protect\ref%
{orbits}.}
  \label{basicG1plot}
\end{figure}

\begin{figure}[htbp] 
  \centering
  \includegraphics[height=6in,keepaspectratio]{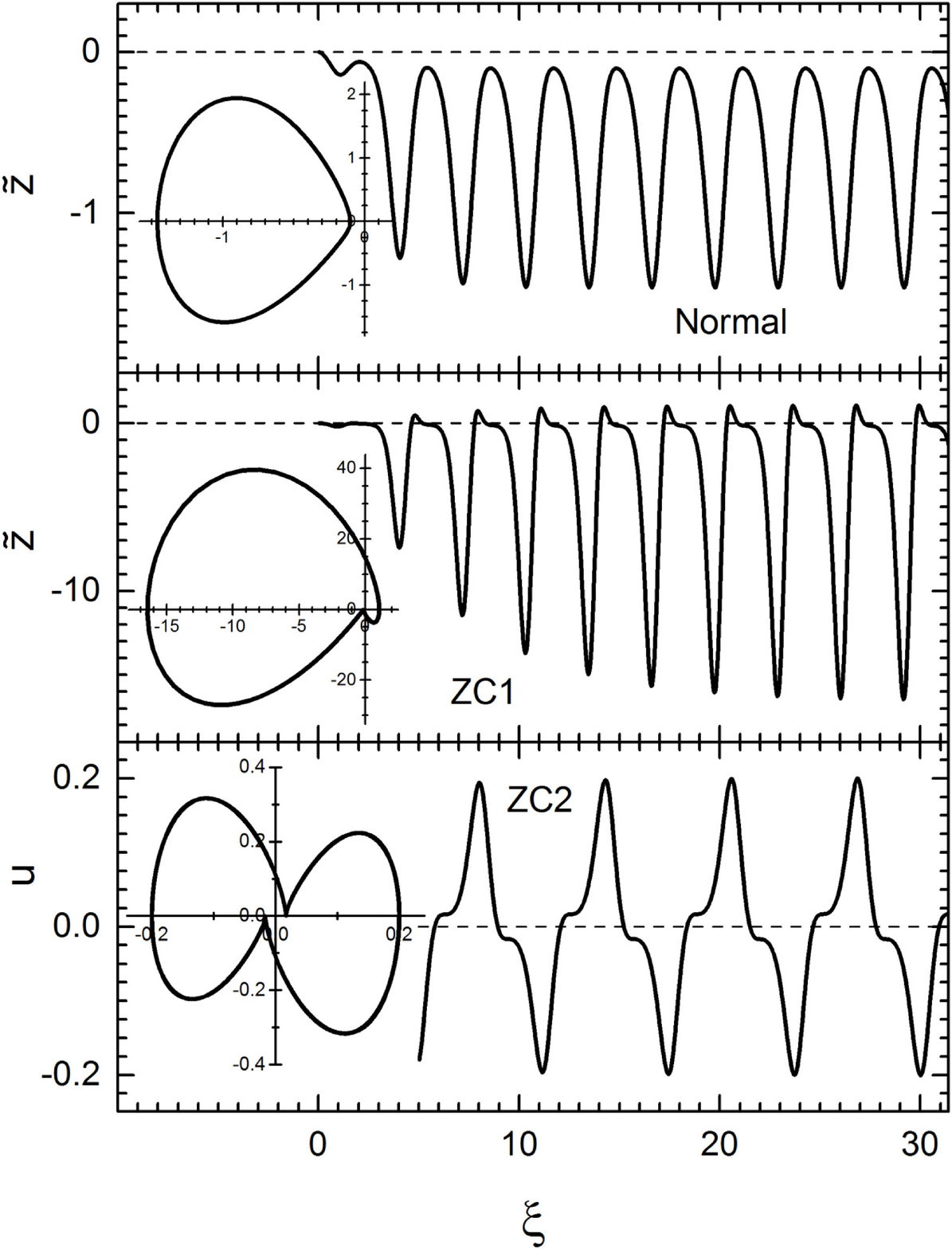}
  \caption{Examples of the Normal (top),
ZC1 (middle), and ZC2 (bottom) extended orbits. The top two panels show $%
\tilde{z}(\protect\xi )$ from solving Equation \protect\ref{withgravity}
(assuming a nonzero $g_{eff})$ starting with the initial conditions $\tilde{z%
}(0)=d\tilde{z}/d\protect\xi (0)=0.$ These two solutions show the particle
dropping down to a stable periodic orbit after a few oscillation periods.
The bottom panel shows $u(\protect\xi )$ from solving Equation \protect\ref%
{NLdampingequation} (which assumes $g_{eff}=0)$ after a stable orbit has
been achieved. Note that the period of the ZC2 orbit is twice that of the
ZC1 orbit. The insets in all three panels show Poincar\'{e} plots of $d%
\tilde{z}/d\protect\xi $ versus $\tilde{z}$ (or $du/d\protect\xi $ versus $%
u) $ after each particle has reached a stable orbit. These three examples
are intended only to show the essential morphologies of the most common
Normal, ZC1, and ZC2 extended orbits; the orbital amplitudes and other
details depend on the specific parameters used in the equations.}
  \label{orbits}
\end{figure}

\begin{figure}[htb] 
  \centering
  \includegraphics[height=3.4in,keepaspectratio]{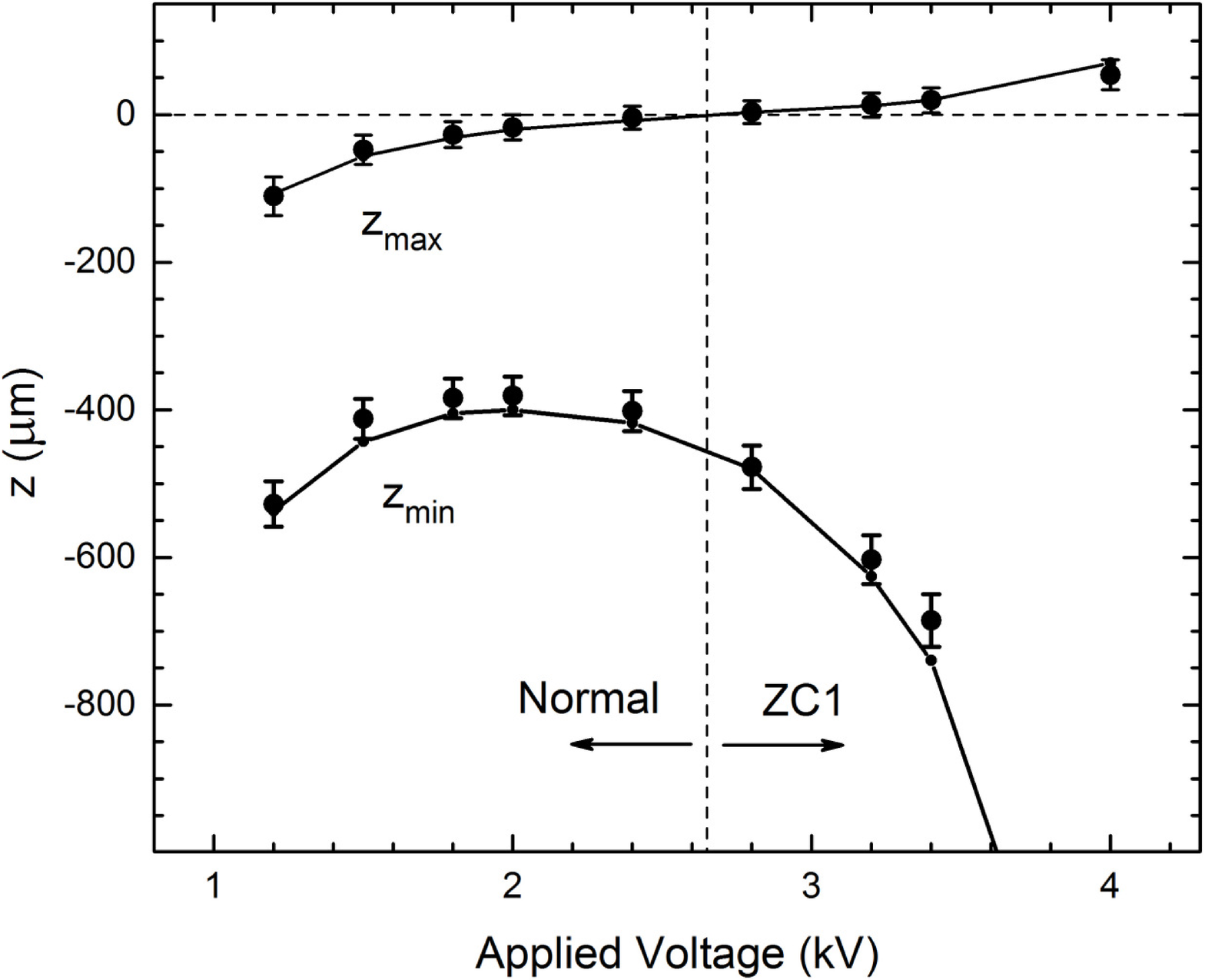}
  \caption{A comparison of measured $%
z_{\min }$ and $z_{\max }$ for a particle in a MQIT (points) with model
calculations (lines). The particle behavior transitioned from Normal on the
left to a ZC1 zero-crossing orbit on the right.}
  \label{ZC1Expt}
\end{figure}

With the additional constant force, there is an accompanying change in the
particle dynamics, which we investigated by integrating Equation \ref%
{withgravity} to directly observe the dynamical behavior of $\tilde{z}(\xi )$
for input $b,$ $q,$ and the initial conditions $\tilde{z}$ and $d\tilde{z}%
/d\xi $ at $\xi =0.$ For this we used \textit{NDSolve} in Mathematica, which
could typically compute 20-30 orbital periods before encountering numerical
instabilities. This was usually sufficient for our purposes, as the
solutions $\tilde{z}(\xi )$ usually settled quickly into stable orbits that
were insensitive to the chosen initial conditions. In difficult cases, we
performed longer integrations using \textit{ode45} in Matlab. By examining
computed particle trajectories with many different inputs, we obtained the
results shown in Figure \ref{basicG1plot}.

When $q$ is above the stability line shown in Figure \ref{basicG1plot}, the
solutions are unbounded and particles are ejected from the trap. Below the
stability line, in the Normal region shown in the figure, the gravitational
force pulls particles down to $\tilde{z}<0,$ where they exhibit simple
oscillatory micromotion. These orbits are stable, as the average trapping
force balances gravity, and an example of this simple motion is shown in the
top panel in Figure \ref{orbits}.

In the ZC1 region in Figure \ref{basicG1plot}, particles exhibit
zero-crossing orbits (i.e. the motion passes through $\tilde{z}=0$), and
again an example is shown in the middle panel of Figure \ref{orbits}. From
Figure \ref{basicG1plot} we see that the ZC1 orbits only occur for $b >
3.5,$ when the particle motion is sufficiently overdamped. In the Normal
region, increasing $q$ pulls the particle closer to the trap center at $%
\tilde{z}=0.$ In the ZC1 region, however, increasing $q$ results in an orbit
with a greater overall extent, which grows to infinity as the stability line
is approached.

Since $\tilde{z}=z\Omega ^{2}/4g_{eff},$ we see that $z\rightarrow 0$ as $%
g_{eff}\rightarrow 0,$ and for $g_{eff}=0$ we confirmed in our numerical
analysis that the stability line in Figure \ref{basicG1plot} separates
stable solutions that decay to $z=0$ from unstable solutions that eject
particles from the trap, This is consistent with the fact that Equation \ref%
{withgravity} reduces to the Mathieu equation for $g_{eff}=0$, and the
stability line in Figure \ref{basicG1plot} is consistent with the related
analysis presented in \cite{foot}.

We confirmed the calculated behavior using a MQIT operating at $\Omega /2\pi
=60$ Hz, loaded with a single Borosilicate glass microsphere having a
specified density of 2230 kg/m$^{3}$. A microscope objective built into the
MQIT allowed us to image the trapped particle directly, yielding a measured
diameter of $2R=9\pm 2$ $\mu $m, which was consistent with the nominal
diameter of $10\pm 1$ $\mu $m specified by the manufacturer. Balancing
gravity with a constant electric field $E_{0}$ gave a measured $%
Q/m=g/E_{0}=0.0176\pm 0.0015$ C/kg. Using Stoke's-law damping in air ($%
\gamma =6\pi \mu R,$ with $\mu =1.8\times 10^{-5}$ kg/m-s) gave $b=9.5\pm 3,$
and the electric field $E_{z}(z,t)$ was calculated from the known geometry
of the trap, giving $q$ as a function of the applied voltage.

With all the relevant particle and trap properties determined, we were able
to create a model of the particle behavior with no free parameters for
comparison with experiment. The uncertainty in the particle radius was
rather large, however, so in the end we adjusted this parameter (within the
stated uncertainty limits) to better fit the data, thus essentially using
the orbital motion to measure $R$. We measured and calculated $z_{\min }$
and $z_{\max }$, the extrema of the particle motion (which was purely axial)
as a function of the applied voltage. Figure \ref{ZC1Expt} shows results
with no bias field to counteract gravity $(E_{0}=0).$ The smooth transition
from Normal to ZC1 behavior was essentially as calculated, and we also
confirmed (not shown in the figure) that the overall scale of the motion was
proportional to $g_{eff}.$

From this and other observations of orbital behaviors in MQITs, we confirmed
the simple 1D theory for a 3D quadrupole trap with linear damping and the
addition of a constant gravity-like force. The Normal and ZC1 orbits were
observed as purely axial motions, and our measurements showed good
quantitative agreement with calculations.

\subsection{Nonlinear Damping}

\begin{figure}[htb] 
  \centering
  \includegraphics[height=3.9in,keepaspectratio]{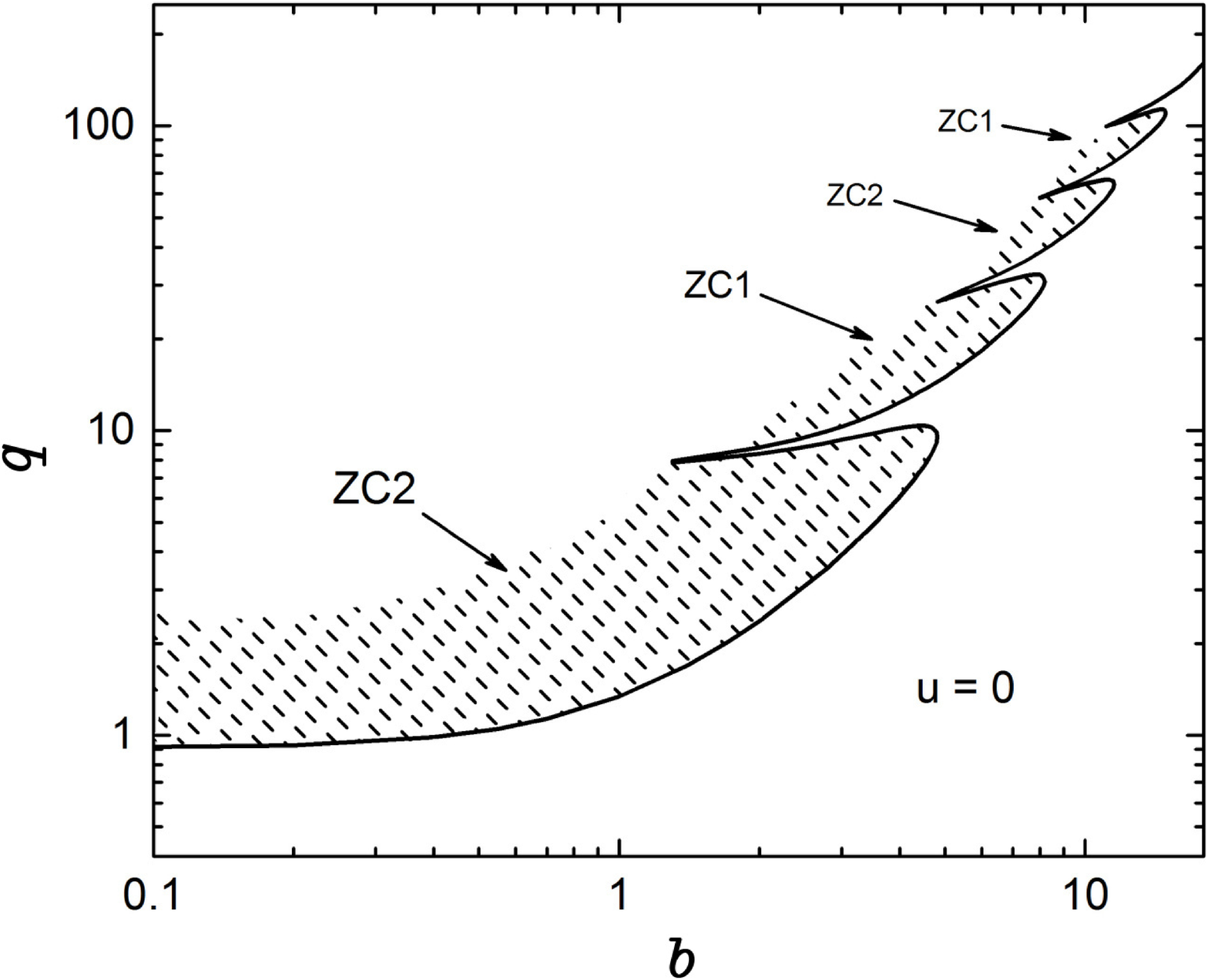}
  \caption{A stability diagram for the
zero-gravity case with $a=0$ and an additional nonlinear damping term,
described by Equation \protect\ref{NLdampingequation}. The stability line is
identical to that shown in Figure \protect\ref{basicG1plot}. Below the
stability line, all solutions decay to $u=0.$ Above the stability line, the
nonlinear damping causes all solutions to be bounded, so no particles are
ejected from the trap. Just above the stability line, particles exhibit
stable ZC1 or ZC2 orbital behaviors that are largely independent of initial
conditions, as labeled in the diagram. Far above the stability line, the
particle behavior is typically aperiodic and strongly dependent on initial
conditions.}
  \label{NLdamping}
\end{figure}

In MQITs, the Reynolds number of the motion is often of order unity or
higher, requiring an additional damping force proportional to $%
v^{2}=(dz/dt)^{2},$ giving the total damping force $-\gamma v\left( 1+\alpha
\left\vert v\right\vert \right) ,$ where $\gamma $ describes the linear
damping and $\alpha $ is a nonlinear damping constant \cite{pruppacher}. For
the zero-gravity case with no DC fields $(a=0),$ the equation of motion then
becomes

\begin{equation}
\frac{d^{2}u}{d\xi ^{2}}+b\frac{du}{d\xi }\left[ 1+\left\vert \frac{du}{d\xi 
}\right\vert \right] -2qu\cos \left( 2\xi \right) =0
\label{NLdampingequation}
\end{equation}%
where $u=\alpha \Omega z/2.$ A dynamical analysis of the solutions to this
equation yields the stability diagram shown in Figure \ref{NLdamping}.

As this is a zero-gravity case, the solutions below the stability line are
all damped to $u=0.$ This makes physical sense, since the additional
nonlinear damping can only increase the stability. Above the stability line,
however, the increased damping is sufficient to prevent the realization of
any unbounded solutions that eject particles from the trap. Just above the
lowest branch of the stability line, particles exhibit a new type of stable
zero-crossing orbit we label ZC2, and an example of this motion is shown in
the third panel in Figure \ref{orbits}. Since the particle spends one AC
cycle above $u=0$ and one AC cycle below, the total period is $4\pi /\Omega
, $ double the period of the AC drive and the other orbital motions
described above, as shown in Figure \ref{orbits}.

For higher $b,$ again just above the stability line, particles exhibit
stable ZC1 or ZC2 orbits as shown in Figure \ref{NLdamping}. Note that the
ZC1 orbits arising from the nonlinear Equation \ref{NLdampingequation} occur
for $g_{eff}=0,$ although their morphology is essentially the same as we
found with the linear $g_{eff}\neq 0$ case described above. Since there is
no effective gravity to break the up/down symmetry in Equation \ref%
{NLdampingequation}, a ZC1 orbit may point up or down depending on initial
conditions. Note also that the boundaries between the ZC1 and ZC2 regions
are not sharp, especially at high $q,$ where the particle motions are
sensitive to initial conditions and may be aperiodic. The ZC1 and ZC2
regions are quite distinct just above the stability line, however,
segregating the two morphologies as shown in Figure \ref{NLdamping}.

The physical size of an extended particle orbit depends on the nonlinear
damping via $z=2u/\alpha \Omega ,$ so as expected $z$ becomes unbounded
above the stability line as $\alpha \rightarrow 0.$ This makes it possible
to measure the nonlinear damping coefficient directly at low Reynolds number
by observing the ZC2 orbital behavior in a MQIT, perhaps to higher accuracy
than has been accomplished to date by more traditional means \cite%
{pruppacher}. For example, for $b<2$ the physical size of a ZC2 orbit scales
approximately as $z\sim 1/\gamma \alpha .$ The linear damping coefficient is
typically Stoke's damping in a MQIT, so the nonlinear damping coefficient $%
\alpha $ can be extracted from a measurement of the orbital size.

\begin{figure}[htb] 
  \centering
  \includegraphics[width=4.25in,height=3.42in,keepaspectratio]{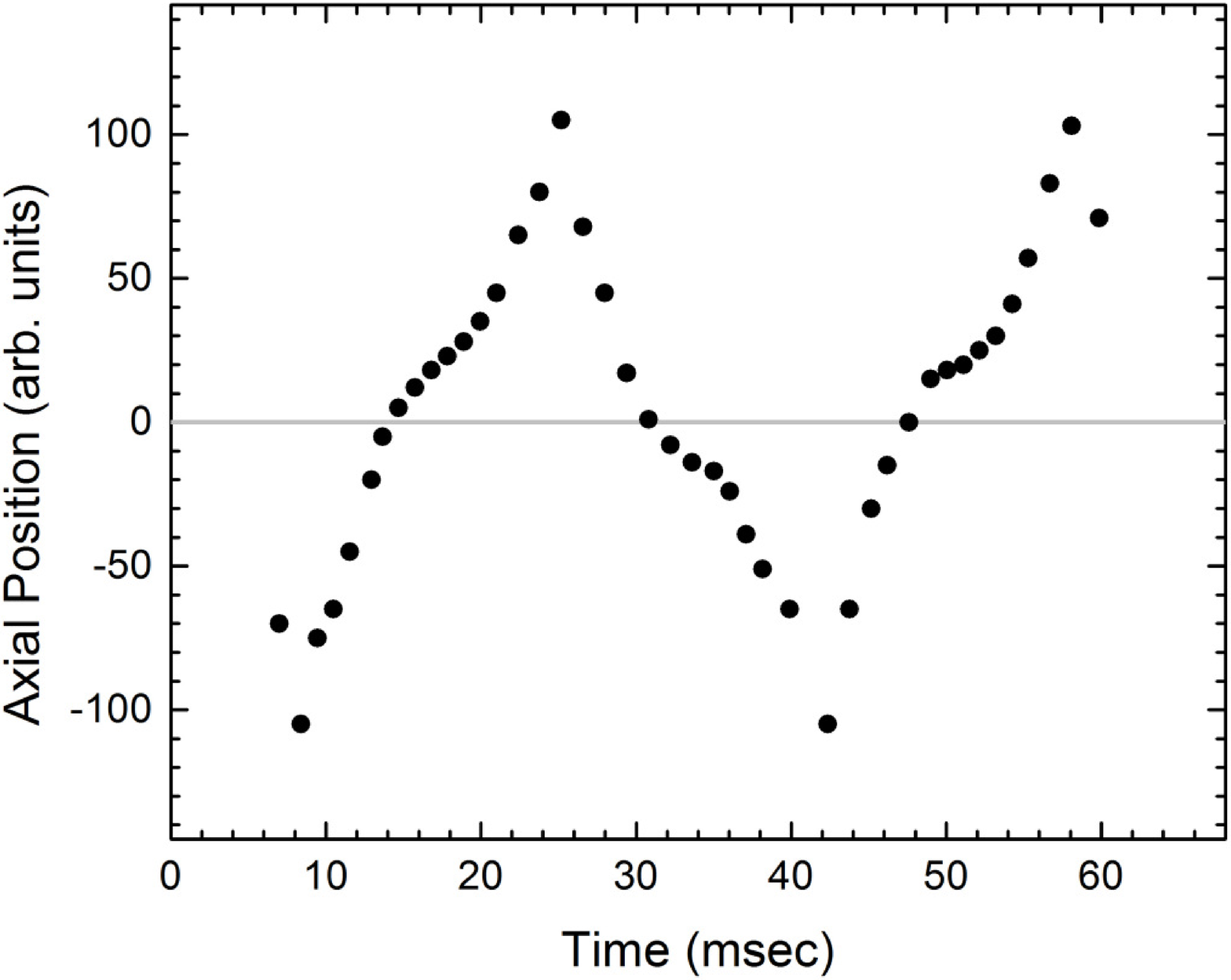}
  \caption{Measurements of a typical ZC2
orbit in a 60 Hz MQIT, showing the axial motion $z(t).$ The radial motion
was confined to $r=0$ in the trap. The zero point of the measured motion
presented in the plot is relative to the trap center, as determined by
cancelling gravity and observing the stable position of the particle at $%
z=0. $ The overall characteristics of the orbital motion are consistent with
that shown in Figure \protect\ref{orbits}, including the 30 Hz oscillation
frequency.}
  \label{ZC2orbit}
\end{figure}

In our MQITs we have found that the ZC2 orbital behavior is quite common,
and Figure \ref{ZC2orbit} shows one example. To obtain this measurement we
strobed the laser illuminating the particle near 30 Hz, allowing the
particle position to be measured from a simple video recording of the
motion. Again the overall characteristics of the observed motion are well
described by the 1D equation of motion. Another example showing ZC2 axial
motion can be found in \cite{vid1}.

\subsection{Nonlinear Electric Fields}

We have also examined particle behavior when the electric field geometry
deviates from the linear $E_{AC}(z)=A_{AC}z$ found in a pure quadrupole ion
trap. There are many simple MQIT geometries in which the electric field
rolls off at high $z,$ and we model these nonlinear field geometries using $%
E_{AC}(z)=k^{-1}A_{AC}\tan ^{-1}(kz),$ where $k$ sets the scale of the
rolloff in the field. With this functional form the field is approximately
linear when $z\ll k^{-1},$ reaching a constant value when $z\gg k^{-1},$ and
taking $k\rightarrow 0$ returns the purely linear field. With this change,
the equation of motion becomes 
\begin{equation}
\frac{d^{2}\tilde{u}}{d\xi ^{2}}+b\frac{d\tilde{u}}{d\xi }-2q\tan ^{-1}(%
\tilde{u})\cos \left( 2\xi \right) =0
\end{equation}%
for the case $a=0$ and $g_{eff}=0,$ where $\tilde{u}=kz.$ Analyzing this
yields a stability diagram quite similar to that shown in Figure \ref%
{NLdamping}, featuring both ZC1 and ZC2 extended orbits. Again the orbits
decay to $\tilde{u}=0$ below the stability line, and there are no unbounded
solutions above this line. Examining other electric field configurations
(for example $E=Ak^{-1}\log (1+kz)$ and $E=Az/(k^{3}z^{3}+1)$), we found
that the stability diagrams were all similar to that shown in Figure \ref%
{NLdamping}, as long as the field rolled off at high $z.$

In \cite{ziaetan2} the authors describe calculated axial ZC2 orbits in a 3D
QIT with linear damping in a purely quadrupole field geometry, in
contradiction to our results. We also found, however, that a ZC2-like
behavior could be seen in the absence of nonlinearities when sufficiently
close to the stability line, and this may explain the discrepancy with \cite%
{ziaetan2}. Integrating the equation of motion for $\sim 20$ orbital periods
(using Mathematica) can yield a ZC2-like orbit that appears to be stable,
but integrating with the same parameters for several hundred periods (using
Matlab) reveals that these orbits are in fact slowly decaying. We only found
truly stable extended orbits with the addition of nonlinear damping or
nonlinear field geometries, or both.

\section{2D Damped Ion Traps}

For the case of a two-dimensional quadrupole field with a nonlinear damping
force $\vec{F}_{damping}=-\gamma \vec{v}\left( 1+\alpha \left\vert \vec{v}%
\right\vert \right) $, the equations of motion in $(x,y)$ coordinates can be
rescaled to become%
\begin{eqnarray}
\frac{d^{2}\hat{x}}{d\xi ^{2}}+b\frac{d\hat{x}}{d\xi }\left[ 1+\left\vert 
\hat{v}\right\vert \right] -2q\hat{x}\cos \left( 2\xi \right) &=&0
\label{2Dquadequations} \\
\frac{d^{2}\hat{y}}{d\xi ^{2}}+b\frac{d\hat{y}}{d\xi }\left[ 1+\left\vert 
\hat{v}\right\vert \right] +2q\hat{y}\cos \left( 2\xi \right) &=&0  \nonumber
\end{eqnarray}%
where $\hat{x}=\alpha \Omega x/2,$ $\hat{y}=\alpha \Omega y/2,$ $b=2\gamma
/m\Omega $, $q=2QA/m\Omega ^{2}$, the AC electric field is $\left(
E_{x},E_{y}\right) =\left( Ax,-Ay\right) ,$ and $\left\vert \hat{v}%
\right\vert =\alpha \left\vert \vec{v}\right\vert =[\left( d\hat{x}/d\xi
\right) ^{2}+\left( d\hat{y}/d\xi \right) ^{2}]^{1/2}.$ Here we have assumed
zero DC electric field and $g_{eff}=0.$ In the absence of nonlinear damping $%
(\alpha =0),$ these equations decouple to the 1D case above, yielding the
same stability line shown in Figure \ref{basicG1plot}. Since we are taking $%
g_{eff}=0$ in this case, our numerical analysis confirmed that particles
either decay to $(x,y)=0$ below the stability line or are ejected from the
trap above the stability line.

\begin{figure}[htb] 
  \centering
  \includegraphics[height=4.0in,keepaspectratio]{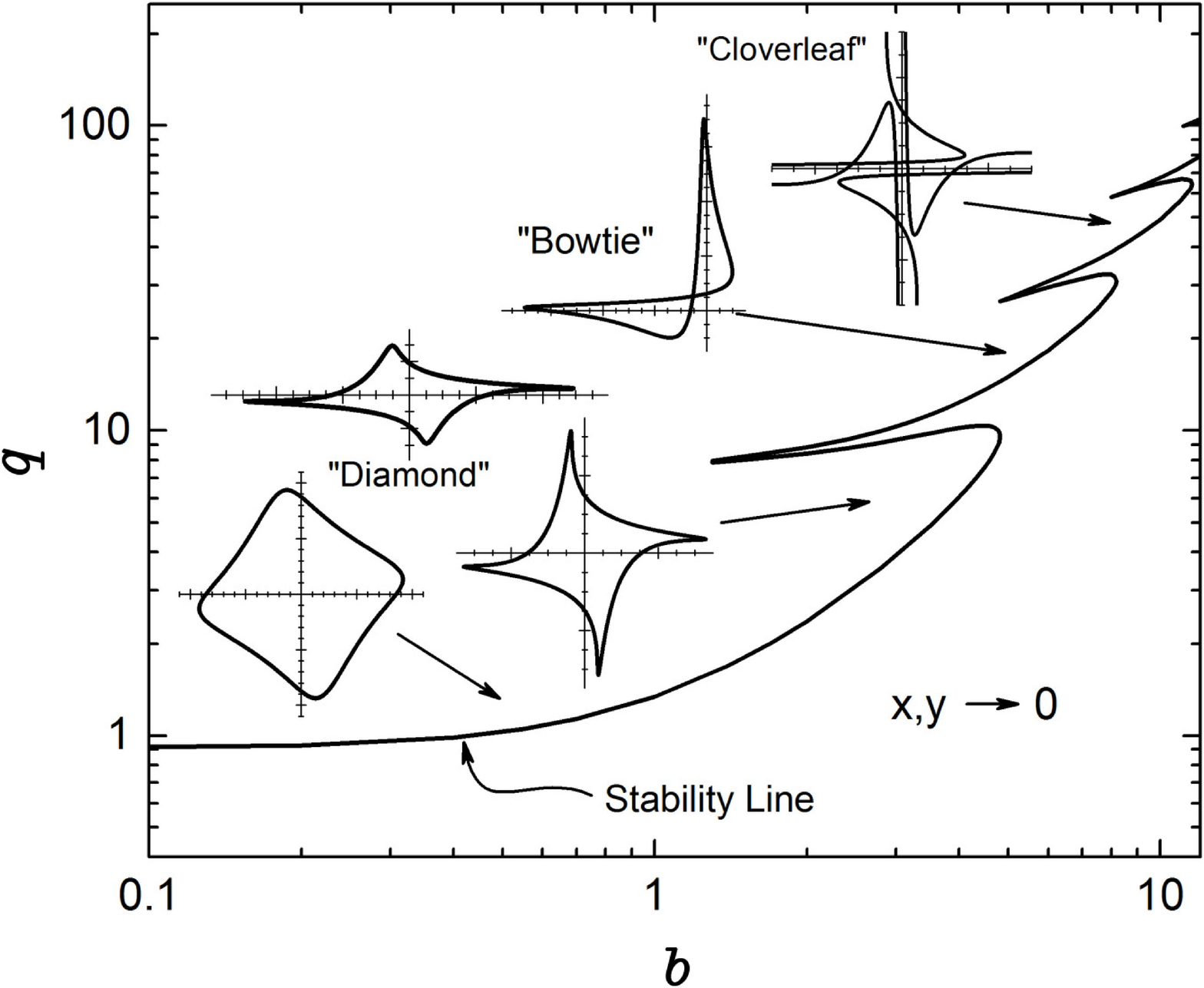}
  \caption{The stability diagram for a 2D
quadrupole trap with nonlinear damping, describing solutions to Equation 
\protect\ref{2Dquadequations}. Below the stability line, particles are
damped to the center of the trap at $(x,y)=0.$ Above the stability line, in
roughly the same regions shown in Figure \protect\ref{NLdamping}, particles
are driven into a variety of stable extended orbits as shown. The inset
plots show examples of calculated closed particle orbits in $(x,y)$ space;
the cloverleaf plot shows only the inner region of the orbit. As with Figure 
\protect\ref{orbits}, the inset diagrams here are intended only to show the
morphologies of the different types of extended orbits; the orbital
amplitudes and other details depend on the specific parameters used in the
equations.}
  \label{2Dstabilitydiagram}
\end{figure}

Including nonlinear damping, however, yields a rich spectrum of stable
extended orbits, as shown in Figure \ref{2Dstabilitydiagram}. The
  diamond-shaped orbits are rounded for $b < 2$ and develop cusp-like
corners at higher $b.$ The orbits are typically more asymmetrical (as shown
in the uppermost example of the three diamond plots) at higher $b$ and
higher $q$, where the symmetry is more sensitive to initial conditions.
Above the second branch of the stability line, particles are driven into
\textquotedblleft bowtie\textquotedblright\ orbits, while more complex
\textquotedblleft cloverleaf\textquotedblright\ orbits appear above the
third branch, as indicated in the figure. These orbits are most stable when $%
q$ is just above the stability line, while at high $q$ the motion can be
aperiodic and strongly dependent on initial conditions.

Note there are many similarities between the 1D and 2D orbits occurring in
3D and 2D quadrupole traps, respectively. For example, a diamond orbit is
essentially a ZC2 orbit in both $x$ and $y,$ while a bowtie orbit is like a
ZC1 orbit in both $x$ and $y.$ The 2D orbits correspond to their 1D analogs
in the stability diagram as well, as can be seen by comparing Figure \ref%
{2Dstabilitydiagram} and Figure \ref{NLdamping}. Similarly, the orbital
period is $2\pi /\Omega $ for the bowties and ZC1 orbits, while it is $4\pi
/\Omega $ for the diamonds, cloverleafs, and ZC2 orbits.

\begin{figure}[htb] 
  \centering
  \includegraphics[width=2.39in,height=3.69in,keepaspectratio]{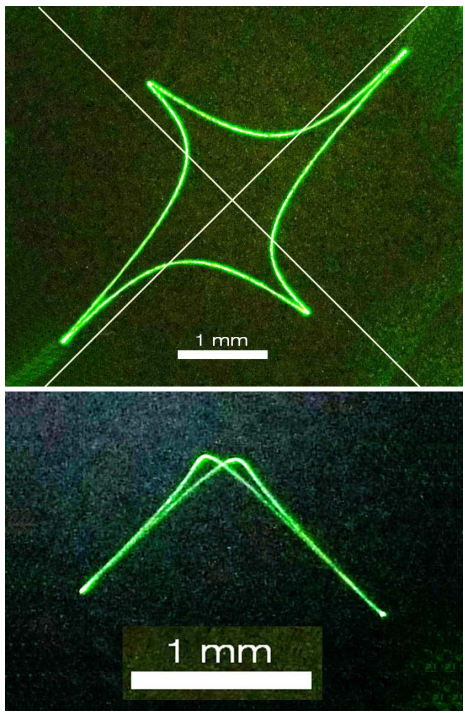}
  \caption{Examples of a diamond (top) and
bowtie (bottom) orbit observed in a 4-bar MQIT with $g_{eff}=0$. The two
photographs show single particles executing stable closed orbits in the $xy$
plane, illuminated by a laser. The exposure times were longer than the
orbital periods, causing the particles to appear as streaks delineating the
orbital paths. Diagonal lines were digital superimposed on the top image to
show the axes of the 2D quadrupole, which had the same orientation for both
images. Scale bars were also added digitally, and a DC bias electric field
canceled the gravitational force.}
  \label{diamondexample}
\end{figure}

Experimentally the diamond orbits are the easiest to obtain, as they require
low damping and a corresponding low drive voltage. The other orbits occur
for larger $b$, and a much larger $q$ is therefore needed to get above the
stability line. To date we have only observed diamond and bowtie orbits in
our 2D MQITs, and Figure \ref{diamondexample} shows two examples. Diamond
orbits can also be seen in the online videos \cite{vid1, vid2}, suggesting
that they are somewhat common in MQITs. We have observed symmetrical and
asymmetrical diamond orbits, along with a variety of odd time-dependent
behaviors we do not yet understand. Diamond orbits with slowly oscillating
changes in the orbital asymmetry were especially common at low $b$, and this
behavior remains puzzling. Some of these behaviors may be driven by residual
air currents within the traps.

The images in Figure \ref{diamondexample} were obtained in a
\textquotedblleft 4-bar\textquotedblright\ MQIT consisting of four identical
conducting bars, collinear in the $z$ direction and arranged on the corners
of a square in the $xy$ plane \cite{coulomb}. The bars had a diameter of 3.2
mm, and their nearest separation was 6.7 mm. An AC voltage was applied to
the bars with alternating polarities to produce approximately quadrupolar
electric fields in $x$ and $y,$ with no electric forces in the $z$
direction. The trap operated at 60 Hz in air, with applied voltages up to 6
kV. Although the electric field geometry in a 4-bar trap is not a pure
quadrupole field, our calculations showed that no extended orbits were
possible in a 4-bar trap in the absence of nonlinear damping. Thus the
behavior of a 4-bar trap is qualitatively the same as a 2D quadrupole trap.

One especially noteworthy phenomenon we can produce on demand in our 2D
MQITs is a collective mode that includes dozens of particles in overlapping
corotating diamond orbits. The phenomenon is difficult to describe
adequately, and still photos show little more than a blur of rapidly moving
trapped particles. Videos are somewhat more informative, and examples can be
found at \cite{newtonianlabs}. Based on our observations of the formation,
behavior, and decay of this collective mode, we believe it is stabilized by
air currents within the trap.

When several particles are initially driven into nearby corotating diamond
orbits, we believe that the particle motions create a fan-like effect that
pushes air radially outward from the trap; that is, the flow is outward in $%
x $ and $y$ away from the trap center at $(x,y)=0.$ This radial outflow is
accompanied by an axial inflow along the $z$ axis. Since there are no
electrical forces in the $z$ direction, the axial inflow pulls the nearby
corotating particles together in spite of the repulsive forces arising from
their like charges. The result is a frenetic knot of particles moving in
overlapping corotating diamond orbits, which we call a \textquotedblleft
trapnado\textquotedblright .

Once a small trapnado forms, the axial airflow quickly pulls in additional
particles that join the knot and reinforce the air motions. Trapnados form
easily with either rotation direction, and we have even witnessed axial
collisions between two independent trapnados \cite{newtonianlabs}, the
outcome depending on their relative rotation directions. We discovered
trapnados serendipitously in our 4-bar MQITs, and we believe that our
hypothesis of air-stabilized diamond orbits provides a sound (albeit
qualitative) physical explanation for this remarkable phenomenon.

\section{Discussion}

Our initial motivation for undertaking this investigation was experimental:
we built a number of MQITs operating in air and soon began seeing extended
orbital behaviors, both 1D axial orbits in 3D MQITs and diamond-shaped
orbits 2D MQITs. Although our online research suggests that others have
observed these behaviors numerous times over the past several decades, our
literature search did not turn up an adequate characterization or
theoretical explanation of the extended orbits.

Our analysis of single-particle trajectories in damped ion traps shows that
nonlinearities are required to produce stable extended orbits, in particular
either nonlinear damping or nonlinearities in the electric field geometry.
In the absence of nonlinearities, particles are either damped to the trap
center or ejected from the trap. As has been documented by others, the
stability line between damped and ejected is sharp, and the only stable
trajectory in this case is the $\vec{x}(t)=0$ solution.

With the addition of nonlinearities, a variety of stable, closed
trajectories appear, as described in detail above. We have not examined all
possible nonlinearities, as this parameter space is large, so additional
novel particle behaviors may exist. Besides 2D quadrupole geometries, we
have also calculated diamond-like orbits in 2D hexapole and octapole
geometries, where the diamonds then show six or eight corners, respectively.
Our focus in the present work was on MQITs operating in air at 60 Hz, driven
by experimental considerations, and we have not yet done an additional
analysis with a focus on atomic/molecular QITs.

Possible applications include measurements of nonlinearity parameters in ion
traps. For example, the orbital motion in a MQIT with nonlinear damping
depends strongly on the nonlinearity parameter $\alpha ,$ somewhat
independently of the linear damping. Because of this, $\alpha $ could be
measured directly even when nonlinear damping is a small perturbation of the
total damping. In contrast, a measurement of a particle's terminal velocity
(for example) yields the total damping only, so $\alpha $ cannot be
extracted independently from such a measurement.

Nonlinearities in the electric field geometry could also be measured using
extended orbital motions, and these measurements are nondestructive in that
particles are not ejected from the trap. Observing the full range of
particle motions both above and below the nominal stability line could allow
independent measurements of many particle properties. Tayloring the electric
field geometry to facilitate a specific measurement of some kind may be
possible.

In addition to calculating single-particle orbits in detail, we also
discovered the remarkable trapnado phenomenon described above. We believe
this is only the second self-organizing collective behavior seen in MQITs to
date, supplementing the well-known Coulomb crystalline structures that have
been observed for many decades.

This work was supported in part by the California Institute of Technology
and a generous donation by Dr. Richard Karp.

\end{document}